*Hydrogen trapping and embrittlement in high-strength Al-alloys*

Huan Zhao[1]*, Poulami Chakraborty[1], Dirk Ponge[1], Tilmann Hickel[1,2], Binhan Sun[1,3], Chun-Hung Wu[1], Baptiste Gault[1,4]*, Dierk Raabe[1]*

[1]Max-Planck-Institut für Eisenforschung, Max-Planck-Straße 1, 40237 Düsseldorf, Germany.
[2]BAM Federal Institute for Materials Research and Testing, Richard-Willstätter-Strβe 11, 12489 Berlin, Germany.
[3]Key Laboratory of Pressure Systems and Safety, Ministry of Education, School of Mechanical and Power Engineering, East China University of Science and Technology, 200237 Shanghai, China.
[4]Department of Materials, Royal School of Mines, Imperial College London, London SW7 2AZ, United Kingdom.

*corresponding authors: H.Z. h.zhao@mpie.de, B.G. b.gault@mpie.de, D.R. d.raabe@mpie.de

## Abstract

Ever more stringent regulations on greenhouse gas emissions from transportation motivate efforts to revisit materials used for vehicles [1]. High-strength Al-alloys often used in aircrafts could help reduce the weight of automobiles, but are susceptible to environmental degradation [2,3]. Hydrogen (H) "embrittlement" is often pointed as the main culprit [4], however, the mechanisms underpinning failure are elusive: atomic-scale analysis of H inside an alloy remains a challenge, and this prevents deploying alloy design strategies to enhance the materials' durability. Here we successfully performed near-atomic scale analysis of H trapped in second-phase particles and at grain boundaries in a high-strength 7xxx Al-alloy. We used these observations to guide atomistic *ab-initio* calculations which show that the co-segregation of alloying elements and H favours grain boundary decohesion, while the strong partitioning of H into the second-phases removes solute H from the matrix, hence preventing H-embrittlement. Our insights further advance the mechanistic understanding of H-assisted embrittlement in Al-alloys, emphasizing the role of H-traps in retarding cracking and guiding new alloy design.

**Main**

High-strength Al-alloys of the 7xxx-series are essential structural materials in aerospace, manufacturing, transportation and mobile communication [5,6], due to their high strength-to-weight ratio, which enables products with lower fuel consumption and environmental impact. The high-strength is achieved through the formation of a high number density (~$10^{24}$ $m^{-3}$) of nanosized precipitates via an aging thermal treatment [6,7]. Unfortunately, high-strength Al-alloys are notoriously sensitive to environmentally-assisted cracking [2,8], and like most high-strength materials, subject to H-embrittlement [9,10]. Overcoming these intrinsic limitations requires gaining a precise understanding of how H penetrates the material and of its interactions with ubiquitous microstructural features, e.g. grain boundaries (GBs) or second-phases, to ultimately cause a catastrophic deterioration of mechanical properties [11]. H-uptake can occur during high-temperature heat treatments, as well as in-service [12,13]. H has low solubility in Al [14], yet crystal defects can assist H absorption [15-22], leading, for instance, to a drop in the fatigue life [23].

The enduring question remains of where the H is located in the microstructure and how such traps facilitate catastrophic failure. Several studies pointed to the critical role of GBs in the environmental degradation. GBs are locations of preferential electrochemical attack [4], but also cracks propagate more easily via GB networks throughout the alloy's microstructure [24,25]. An experimental validation of the H-distribution in Al-alloys is challenging due to its low solubility and to the experimental difficulty of performing spatially-resolved characterization of H at relevant scales and at specific microstructural features. Recent efforts in atomic-scale H-imaging in steels led to insights into the trapping behaviour of second-phases and interfaces [26-28].

**Results and discussions**

Here, we address these critical questions by using the latest developments in cryo-atom probe tomography (APT) [26-28], enabled by cryo-plasma focused-ion beam (PFIB) specimen preparation to investigate H associated with different microstructures in an Al-alloy. Through isotope-labelling with deuterium (D), we partly avoid characterization artifacts associated to the introduction of H from the sample preparation [28,29] and from residual gas in the atom probe vacuum chamber. We studied a 7xxx Al-alloy with a composition of Al-6.22Zn-2.46Mg-2.13Cu-0.155Zr (wt.%) in its peak-aged condition. For this alloy, electrochemical-charging with H leads to a critical drop in the ductility compared with uncharged samples (Fig. 1a). The H desorption spectra are shown in Extended Data Fig. 1. Fig. 1b-d highlights the complexity of the microstructure across multiple length scales. First, Fig. 1b-c reveals the predominant role of GBs and GB networks in the crack formation and propagation during deformation of the H-charged material. Fig. 1d displays the typical distribution of fine precipitates in the bulk, coarse precipitates at GBs and precipitate free zones (PFZs) adjacent to GBs. Intermetallic phases (e.g. S-phase) and $Al_3Zr$-dispersoids that act as grain refiners are also shown.

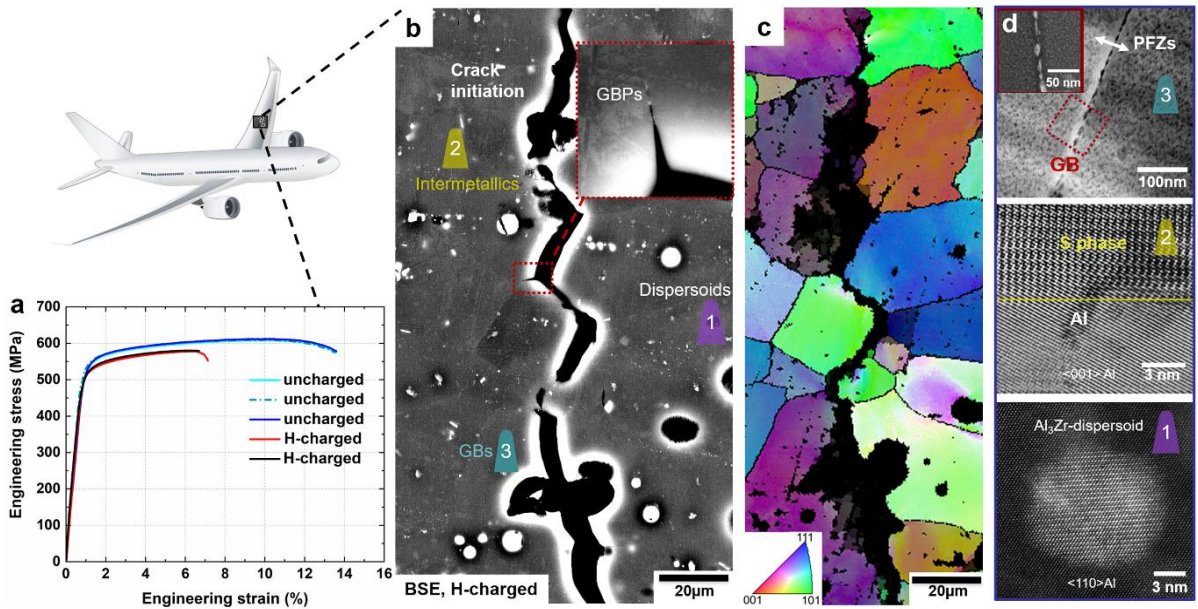

*Fig. 1. Heterogeneous microstructure of an aerospace Al-Zn-Mg-Cu alloy.* (a) Engineering stress-strain curves of uncharged and H-charged samples in the peak-aged condition (120°C/24h). (b) Backscattered electron imaging of an intergranular crack of the H-charged alloy subjected to tensile fracture. (c) Electron backscatter diffraction imaging showing the crack along GBs. (d) The microstructure of GBs, precipitates, PFZs and main types of secondary phases. The color schemes reflect the microstructures where specific APT analyses were performed. GBPs: grain boundary precipitates; GB: grain boundary; PFZs: precipitate free zones; APT: atom probe tomography.

Peak-aged specimens were electrochemically charged with D for subsequent APT probing after validating that H and D show a similar embrittling effect on mechanical properties (Extended Data Fig. 2). D-charged specimens were prepared by PFIB at cryogenic temperatures to limit the introduction of H [29], and immediately analyzed by APT using voltage pulsing to minimize residual H from APT [28,29]. Fig. 2a displays the APT analysis of $Al_3Zr$-dispersoids in the D-charged specimen, with the corresponding composition profile shown in Fig. 2b. H is strongly enriched, up to 9.5 at.% on average, within the dispersoids, contrasting with the much lower content of only 0.4 at.% H in the Al-matrix. An enrichment of D is also revealed with 2.8 at.% inside the dispersoids. H and D atoms partition preferably to sites inside the dispersoids, with a slightly higher content at the interface that may be due to the misfit strain (0.75%) [30]. We further analyzed uncharged specimens prepared by PFIB and electrochemical polishing for reference (Extended Data Fig. 3). H appears consistently enriched inside $Al_3Zr$- dispersoids, up to 8.5 at.% on average. Only a peak at a mass-to-charge ratio at 1 Da, corresponding to H$^+$ atomic ions, is detected in the dispersoids in uncharged specimens. However, in the D-charged material, a distinct peak at 2 Da gives proof of efficient D-charging, with D partitioning into Al$_3$Zr-dispersoids.

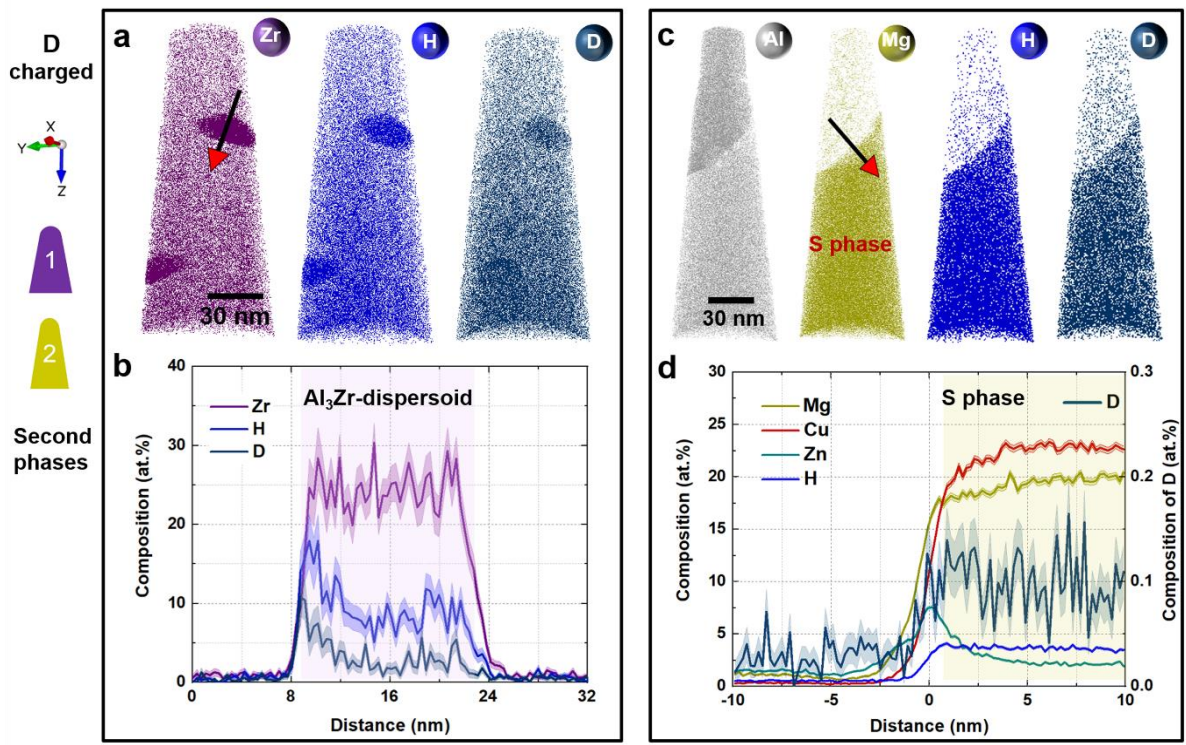

*Fig. 2. APT analysis on second-phases of the D-charged Al-Zn-Mg-Cu samples in the peak-aged condition (120°C/24h). Atom map and composition profiles are presented along the red arrows respectively for (a,b) Al$_3$Zr-dispersoids and (c,d) S-phase. The shaded regions in the profile corresponds to the standard deviations of the counting statistics in each bin of the profile; APT: atom probe tomography.*

Figure 2c shows the APT analysis on an intermetallic particle in the D-charged sample. The composition profile indicates that the Mg-enriched region corresponds to the S-phase (Al$_2$CuMg). The S-particle contains 4.2 at.% H, whereas the matrix has only 0.3 at.% H, and 0.12 at.% D (right axis). Comparison with a similar S-particle in an uncharged sample (Extended Data Fig. 4) shows a 6.5 times higher peak ratio of 2Da/1Da in the D-charged sample, revealing that most of the signal at 2 Da comes from electrochemically charged D. Further evidence of an enrichment up to 9 at.% H within $Al_7Cu_2Fe$, and T-phase particles is provided for the uncharged material (Supplementary Fig. 1-2).

We then analyzed the distribution of H and D at a high-angle GB. Following sharpening at cryo-temperature, the specimen was transferred through a cryo-suitcase into the APT to minimize out-diffusion of D [28]. The peak-aged sample contains 5 nm-(Mg,Zn)-rich strengthening precipitates in the bulk and coarser 20–50 nm-sized precipitates at the GB [31], as well as typical PFZs adjacent to the GB (Fig. 3a). Atom maps of H and $D(H_2^+)$ in Fig. 3b reveal a higher concentration at the GB. Fig. 3c shows details of the precipitates and solutes populating the GB. $Al_3Zr$-dispersoids at the GB (Fig. 3d) contain 11 at.% H and 0.6 at.% D, i.e. a lower D content compared to the $Al_3Zr$-dispersoids in the bulk (Fig. 2b). No enrichment in H and $D(H_2^+)$ (right axis) is shown in (Mg,Zn)-rich precipitates distributed both at the GB (Fig. 3e) and in the bulk (Extended Data Fig. 5). Fig. 3F gives a composition profile through the GB between the particles, showing that the GB is enriched with 2 at.% Mg. We observe no enrichment in Zn and Cu (1 at.%, Extended Data Fig. 6), and in the peak-aged state this can be explained by the accelerated GB precipitation through the consumption of segregated solutes [31]. The locally increased content of $D(H_2^+)$ implies that the solute-decorated GB (i.e. devoid of precipitates) acts as a trap for H, while no enrichment in H and D is observed in the adjacent PFZs (i.e. regions next to the GB), an effect which amplifies the mechanical and electrochemical contrast in these regions. Comparison with a similar GB in an uncharged sample (Extended Data Fig. 7) shows a higher signal at 2 Da (by a factor of 3) in the D-charged sample, supporting that D is indeed enriched at the GB. We obtained 7 APT datasets containing GBs in D-charged samples, and all show consistent enrichment of H and D at GBs (2 additional datasets are shown in Supplementary Fig. 3-4).

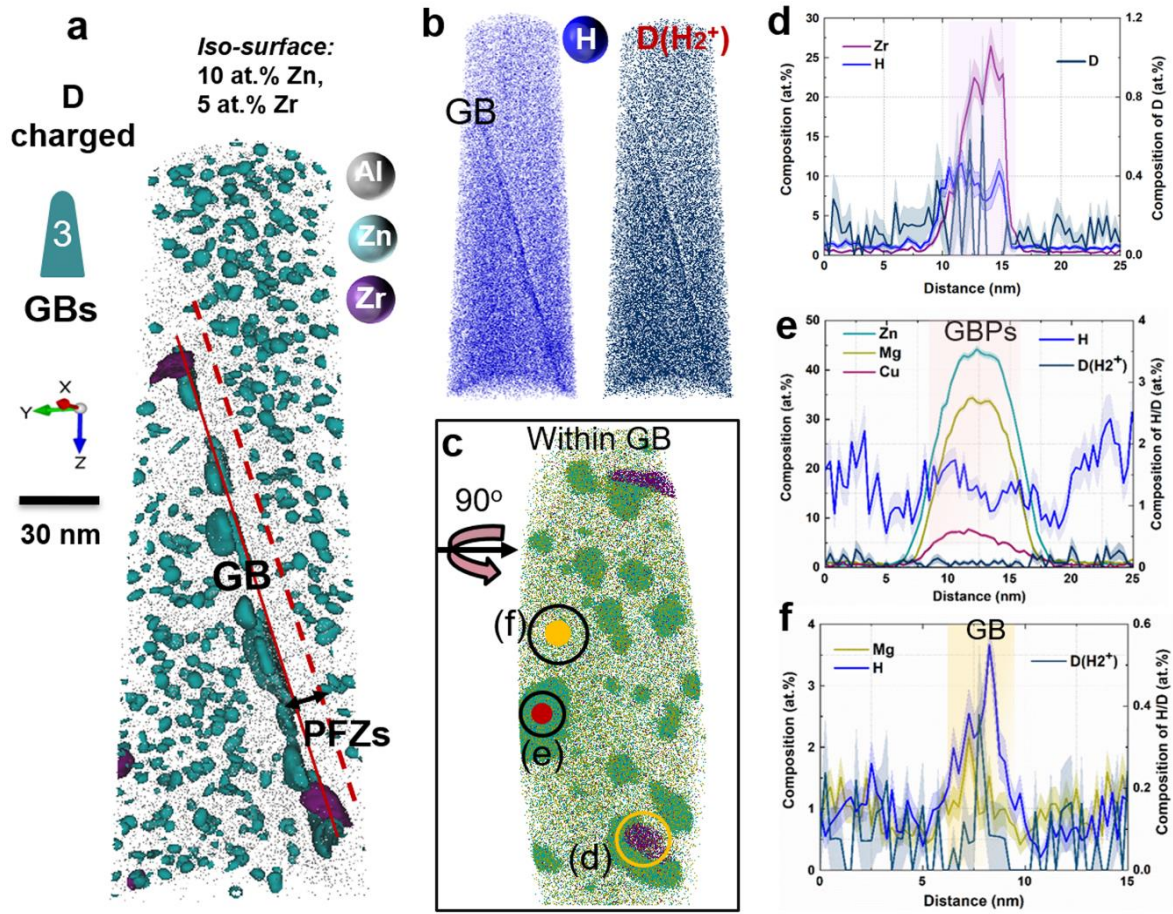

*Fig. 3. APT analysis of a D-charged peak-aged Al-Zn-Mg-Cu sample containing a GB (120°C/24h).*
*(a) The isosurfaces highlight the dispersion of fine (Mg,Zn)-rich precipitates in the matrix, coarse ones at the GB, and Al$_3$Zr-dispersoids. (b) Atom maps of H and D(H$_2^+$). (c) Solute distribution at the GB plane. (d) Composition profile across one Al$_3$Zr-dispersoid at the GB. (e) Composition profile of one (Mg,Zn)-rich precipitate at the GB. (f) Solute composition profile across the GB in between of precipitates. GBPs: grain boundary precipitates; GB: grain boundary; PFZs: precipitate free zones.*

Note that the probability to detect spurious H from residual gas in APT decreases as the strength of the electric field increases, which can be traced by the evolution of the charge-state ratio of Al (i.e. Al$^{2+}$/Al$^{1+}$) [32]. For each microstructural feature studied herein, this ratio is reported in Extended Data Fig. 8, and in each H-enriched case, the electric field either does not change notably or increases compared to Al-matrix. The content of H and D measured in each feature in the uncharged and D-charged conditions is compiled in Supplementary Table 1. These analyses indicate that the peak at 2 Da is extremely unlikely to be associated to H$_2^+$, but to D in D-charged samples, and that most of the detected H was from initially trapped atoms inside the specimen, either from its preparation or/and from the material's processing history [28]. The electrolytical-charging with D reinforces our observation that H is trapped within the material itself [28].

To better understand the effect of H in the intermetallic phases (e.g. S-phase $Al_2CuMg$), $Al_3Zr$-dispersoids and at GBs, we used density functional theory (DFT). Solubility analysis of H in the S-phase reveals that Al-rich octahedral sites provide the lowest solution enthalpy (0.014 eV). The calculated concentrations of H in these sites is 3 at.% at 300 K, substantially higher than $5\times10^{-5}$ at.% assumed for the Al-matrix which explains the APT observations. In $Al_3Zr$-dispersoids, H prefers octahedral interstitial sites with Zr in the second nearest-neighbour (NN) shell with a solution enthalpy of 0.128 eV and a H solubility of 0.2 at.%. However, the high experimental H concentrations may be explained by the presence of Zr-antisites in the first NN positions of H, which reduces the solution enthalpy to -0.202 eV. The solubility of H in the GB was estimated for a symmetric Σ5 (210) [100] GB (Fig. 4a) as a representative high-angle GB [33]. The excess volume for all considered GB sites (Fig. 4b) explains the negative segregation energies given in Fig. 4c. Therefore, the corresponding solution enthalpies at these sites are lower than in the Al-matrix, but still much higher than in the S-phase or $Al_3Zr$-dispersoids.

To explain why GBs, nevertheless, show higher susceptibility for H-embrittlement, as documented in Fig. 1, we determine the embrittling energy associated with H (Fig. 4c). This quantity describes the thermodynamic driving force for fracture formation by comparing the impact of H on the energetics of the GB with that of the free surface (FS). In the Σ5 GB, H when located at sites with the strongest segregation energy, also yields the strongest embrittlement. When distributing H atoms according to their nominal solubility over all these possible sites in a unit area of the GB, weighted by their respective segregation energy (Fig. 4c), the total contribution to the embrittling energy adds up to 600 mJ/m$^2$ for this GB. This value is substantially more positive (i.e., more detrimental) than the embrittling energy determined for Al$_3$Zr-dispersoids (2 mJ/m$^2$) and the S-phase (129 mJ/m$^2$).

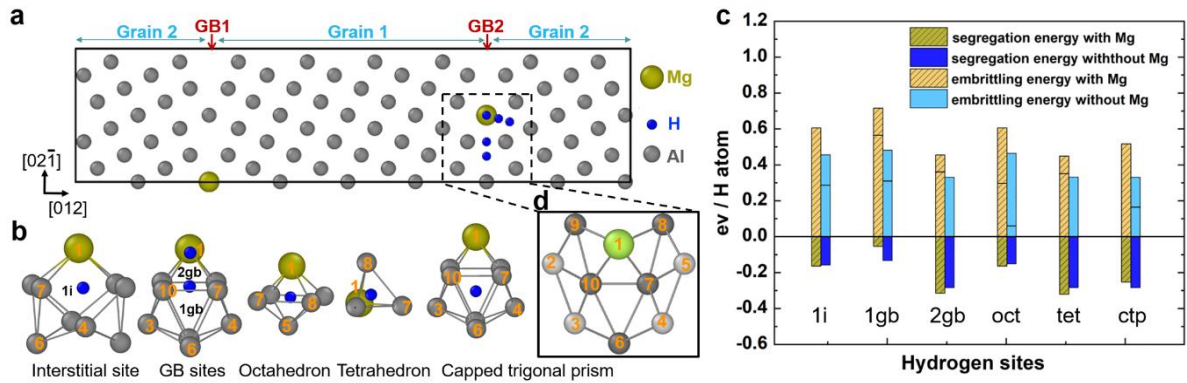

*Fig. 4. Theoretical analysis based on DFT simulations. (a)* Schematic representation of the symmetric Σ5 (210) GB in Al shown with two GB planes. *(b)* The projected and perspective views of deltahedral packing units show the H adsorption sites of the calculations. Site number 1 is the substitutional site for a Mg atom nearby the H sites located inside the polyhedral packing units. *(c)* The embrittling energy and segregation energy are compared in absence and presence (patterned bars) of Mg as a solute atom at the GB for different interstitial sites of H. *(d)* The Al (light grey) and Mg atoms (light green) in the enlarged figure belong to different adjacent (001) planes. DFT: density functional theory.

We investigate the impact of Mg segregation to GBs revealed by APT and introduce in the simulation cell a Mg atom substituting one of four equivalent Al atoms in the GB plane (Fig. 4d). The negative segregation energy of Mg (-0.274 eV) indicates that it stabilizes the GB compared to defect-free Al [34,35], whereas the small negative embrittling energy (-0.043 eV) yields almost no effect on the GB strength compared to the formation of free surfaces. However, for H added to the GB supercell into the interstitial sites at, and near the segregated Mg atom (Fig. 4a), the embrittling energy changes dramatically as summarised in Fig. 4c. The solution enthalpy gives no indication that co-segregation of Mg and H is energetically favorable. In particular, H sitting at the capped trigonal prisms maintains its strong (i.e. negative) segregation energy and has a strong (i.e. positive) embrittling energy which is significantly enhanced in the presence of Mg. In the same way, all other sites substantially contribute to embrittlement when a Mg atom is nearby and when H diffusion at the opening free surface is considered. This is even true for sites like *1i* and *1gb*, for which an occupation by H is less likely. Overall, these effects increase the embrittling energy by H per unit GB area by approximately one order of magnitude with Mg compared to the Mg-free case. The opposite impact of Mg on the segregation and the embrittlement caused by H is explained by the interaction of Mg and H at the free surface resulting from the decohesion.

We can now rationalize the H-embrittlement mechanism as follows: As H penetrates the alloy, the equilibrium H concentration remains low in the Al-matrix and also in the fine and coarse (Mg,Zn)-rich precipitates. However, H accumulates in intermetallic phases (e.g. S or T phase), $Al_3Zr$-dispersoids, and to a lesser extent, at GBs. The high H-enrichment in the second-phase particles was explained by DFT calculations where, H shows no clear decohesion or embrittlement effects. Upon H saturation of the second-phases, H when, intruding further, will gradually accumulate at GBs. DFT predicts no dramatic increase in H concentrations in presence of Mg, which agrees with APT where H is not strongly segregated at GBs compared to second-phases. Yet DFT calculations suggest that when combined with Mg, the strong driving force for H to segregate to a free surface with respect to a possible interstitial site at GBs favors GB decohesion and drives the system towards crack formation. This rationalizes that GBs are embrittled and explains that Mg can impact the H-embrittlement without promoting the absorption of H to GBs [11,36]. Further investigation on the elemental distribution at a H-induced intergranular crack using scanning Auger electron microscopy reveals the enrichment of Mg at the cracked GB (Extended Data Fig. 9). The enrichment is even stronger (by a factor of 2) than the Mg concentration at the GB (Fig. 3f). To confirm the generality of the role of Mg we also show that no H-embrittlement features occurred in a Mg-free Al-5.41 wt.% Zn alloy that was used as reference material. The alloy was exposed to the same H-charging and tensile test conditions, but no sign of H-embrittlement was found, neither in the tensile test results nor in the metallographic fractography (Extended Data Fig. 10). These findings support the result that the co-segregation of Mg and H to free surfaces provides the driving force for the embrittlement of GBs.

**Conclusions**

Generally, avoiding the ingress of H in the first place is extremely unlikely to work, and the best approach to mitigate H-embrittlement is therefore to control its trapping to maximise the components' lifetime in-service. Our results provide indications of H-trapping sites and their respective propensity to initiate damage in the environmentally-assisted degradation, thus contributing to establish a mechanistic understanding of H-embrittlement in Al-alloys. Which specific measures can be explored on the basis of this study, to enhance the resistance to H-induced damage, thus improving the lifetime and sustainability of high-strength lightweight engineering components? The results on the high H-enrichment in second-phase particles provide a potential mitigation strategy for improving H-embrittlement resistance, namely, through introduction and manipulation of the volume fraction, dispersion, and chemical composition of the second-phases, despite their potentially harmful effects on mechanical properties. Other strategies could aim at designing and controlling GB segregation, for instance with the goal of eliminating Mg decoration of GBs by trapping it into precipitates and keeping it in bulk solution. A third and more general avenue against environmental degradation lies in reducing the size of precipitation free zones in these alloys, with the goal to mitigate the H-enhanced contrast in mechanical and electrochemical response between the H-decorated GBs and the less H-affected adjacent regions.

**Acknowledgment**

We gratefully acknowledge Mr. Andreas Sturm for the technical support with cryo-experiments in the PFIB and cryo-suitcase transfer in the atom probe. The help of Dr. Leigh Stephenson for the cryo-transfer in the atom probe is also appreciated. We are grateful to Dr. Di Wan for the initial TDS measurements, and Mr. Michael Adamek for the tensile experiments. B.G. acknowledges financial support from the ERC-CoG-SHINE-771602.

**Author contributions:** H.Z., B.G., D.R., and D.P. developed the research concept; H.Z. was the lead experimental scientist of the study; H.Z., B.G., and D.R. discussed and interpreted the APT results; P.C. and T.H. performed atomic calculations; B.S. conducted TDS measurements; C.W. performed scanning Auger mapping measurements; H.Z., B.G., P.C., and T.H. wrote the manuscript. All authors contributed to the discussion of the results and commented on the manuscript.

**Competing interests:** No

**Data availability:** All data to evaluate the conclusions are present in the manuscript and the supplementary materials.